# INTERPOOL
# A LIQUIDITY POOL DESIGNED FOR INTEROPERABILITY
# THAT MINTS, EXCHANGES, AND BURNS

## HENRIQUE DE CARVALHO VIDEIRA

henriquevideira@live.com | henrique.videira@bcb.gov.br | henrique.carvalho@coppead.ufrj.br
https://orcid.org/0000-0001-9362-2244

Brazilian Central Bank | Rio de Janeiro – RJ | www.bcb.gov.br
Digital Brazilian Real Initiative (DREX CBDC)
Federal University of Rio de Janeiro (UFRJ) | www.ufrj.br
Ph.D. in Finance | Interoperability Research

## ABSTRACT

The lack of proper interoperability poses a significant challenge in leveraging use cases within the blockchain industry. Unlike typical solutions that rely on third parties such as oracles and witnesses, the interpool design operates as a standalone solution that mints, exchanges, and burns (MEB) within the same liquidity pool. This MEB approach ensures that minting is backed by the locked capital supplied by liquidity providers. During the exchange process, the order of transactions in the mempool is optimized to maximize returns, effectively transforming the front-running issue into a solution that forges an external blockchain hash. This forged hash enables a novel protocol, Listrack (Listen and Track), which ensures that ultimate liquidity is always enforced through a solid burning procedure, strengthening a trustless design. Supported by Listrack, atomic swaps become feasible even outside the interpool, thereby enhancing the current design into a comprehensive interoperability solution.

# 1. INTRODUCTION

Full interoperability between distinct blockchains can unlock several use cases in the decentralized realm of blockchains, such as seamless atomic swaps between different cryptocurrencies. However, the current issues in interoperability addressed in this article, such as relying extensively on third parties, still lock away a world of tokenized assets (Heines et al. 2021). In addition to turning assets into interoperable tokens, such tokens are only tradeable where there exists a liquid environment for their exchange. Thus, under these assumptions, how should such interoperability enable liquid tokens without relying on third parties? Before addressing this question in interpool, which is the solution designed in this article, the state-of-the-art solutions in industry are discussed.

The leading interoperability solutions in the blockchain industry are cross-chain bridges, relay chains, decentralized oracles, and wrapping methods. There are several types of wrapped tokens, ranging from centralized approaches, such as Wrapped Bitcoin (Kyber et al. (2019)), to decentralized versions like Staked Bitcoin (Stacks (2022)). Wrapping, in general, is a procedure that mints a token in the mainnet by mirroring such token at a 1:1 parity ratio with an external cryptocurrency; on the other hand, the external cryptocurrency is frozen to guarantee the parity ratio, being released only when a user burns the token at some point. This wrapping can either rely on a centralized party to transfer frozen external coins (e.g., Wrapped Bitcoin) or can be managed by several decentralized parties that sign together a single vault where the frozen assets are kept; the signature of this single vault employs a threshold signature (Schnorr (1990)) in most cases. A technical drawback in decentralized approaches is that part of the stakers set who sign the threshold wallet can be eventually offline, triggering recovery modes. Another issue related to this approach is that the threshold signers usually must keep a 200% ratio of native coins over the external coin (alien coin), which leaves an excess of capital without yields. In this article, the term alien corresponds to coins and smart contracts that are located outside the mainnet in another blockchain. A complete glossary of terms is located before the References Section.

Other state-of-the-art approaches for interoperability are cross-chain bridges provided by companies such as Axelar (Axelar (2021)). The cross-chain bridges are blockchains whose nodes listen to the state of alien blockchains. In order to listen to alien chains, the nodes are required to run full clients to confirm messages and transactions. The transactions are commonly signed by a set of validators using a threshold signing schema, such as Schnorr. On the side of relay chains, Polkadot (Wood (2016)) was conceived with the interoperability concern, allowing seamless transactions between parachains (satellite layers) and a central core (relay chain) that validates every transaction. Even though Polkadot is an interoperability solution on its own, the design requires specific bridges to other blockchains, such as Bitcoin (Nakamoto (2008)) and Ethereum (Buterin (2014)). From another perspective, Chainlink (Ellis et al. (2017)) is a decentralized oracle network that pushes requested data to the mainnet, where an aggregating contract computes a weighted answer to the requester. A comparison between interpool and other industry approaches is shown in Figure 1.

The interpool design arose from the intuitive reasoning that interoperability should enable not just seamless exchanges but also provide liquid environments for flash swaps. However, there are already some solutions in the market that provide flash swaps, such as liquidity pools and typical exchanges like Coinbase, Kraken, and Binance. While typical exchanges are central parties subject to the inherent risk of poor governance, liquidity pools present other issues for the investors, such as the risk of not redeeming wrapped tokens. As liquidity pools are not responsible for wrapped tokens, the risk of not redeeming a token relies on an external entity. Unlike typical liquidity pools, interpool is the sole responsible for minting and burning tokens that are exchanged in its own environment. Besides minting and burning, interpool also acts as a liquidity pool that exchanges its own freshly minted tokens by mainnet coins (native coins). Therefore, interpool is a combination of a liquidity pool and a wrapping system, comprising in just one place minting, exchanging and burning (MEB).

| INTERPOOL | AXELAR | POLKADOT | sBTC |
|---|---|---|---|
| **MAIN FEATURES** | | | |
| A combination of a liquidity pool and a wrapping system, where minting, exchanging and burning (MEB) occur in one place. In interpool design, an alien blockchain hash is forged by optimizers, and this hash supports a new protocol named Listrack that ensures trustless burning transactions. The forged hash and Listrack are the foundations that enable the MEB approach in a single pool without requiring external wrapped coins. | A blockchain specifically designed for interoperability is supported by a non-permissioned set of nodes that run full clients from other blockchains. The consensus based on a delegated proof-of-stake ensures executing transactions signed by a threshold schema. | It allows the connection of any parachain created with Substrate to any other parachain, enablin full interoperability within the same ecosystem. Security is enforced by consensus in the relay chain, which is the core of the system. | Wrapped Bitcoins are kept in a vault controlled by a set of threshold signatures. It runs on top of Bitcoin (layer 2). The set of signers is dynamically changed by a rotating number of staking validators. . |
| **STRENGHTS** | | | |
| It does not rely on third parties to either transfer or back a token, supporting MEB in just one place. Other actors, such as miners, boosters, and liquidity providers, can benefit from greater fees derived from a greater volume of transactions due to optimization. Interpool is a comprehensive interoperability approach that supports atomic swaps even for other smart contracts located outside interpool. | It is blockchain agnostic. The design intends to create an ecosystem where developers can code only once, delivering their DeFi applications for several blockchains without worrying about different languages and data structures. | Full interoperability within its own ecosystems, comprised of multiple connected parachains. | Unlike typical centralized wrapping, such as wBTC, sBTC does not rely on a centralized entity. |
| **DRAWBACKS** | | | |
| The logic depends on a large number of transactions, but such issue can be mitigated by using typical wrapped tokens to initiate interpool activity until reaching a minimum threshold. A front-end application must be developed for liquidity providers in order to listen for critical events, such as burning and collateral injections. | Security relies on the nodes' honest behavior, which plays a vital role in minting, burning and transferring. Possible stalls can result in locking external assets forever, though mitigation procedures exist. | Bridges with other blockchains still depend on specific designs, such as Xclaim (Bitcoin) and Hyperbridge (Ethereum). | It relies extensively on external data provided by witnesses, such as the exchange ratio and whether transactions were processed. |

Figure 1: Comparison between state-of-the-art solutions and interpool

In interpool, the liquidity pool becomes a new type of pool where the order of transactions is optimized, which is a novelty in the defi environment. In typical liquidity pools, such as Uniswap (Hayden et al (2020)), the order of transactions that reach the smart contract is defined by the miner of the block. Regularly, a miner sorts out transactions based on their gas prices, which can lead to a front-running misbehavior. The front-running happens when a bad actor targets a large transaction in the mempool and places another one with a higher gas price just before it, generating an effect that gives him higher profits. A special type of front-running is the sandwich misbehavior, which happens in liquidity pools when bad actors dig the greatest volume transaction. After finding this target transaction, they place two specific orders that are located just before and after the target transaction. As the target transaction significantly impacts the exchange ratio, the bad actors profit from the price difference by buying the coin at a lower price (their first injected order) and selling it at a higher price (their second injected order).

The front-running issue is solved in interpool because the new design takes advantage of the order book issue to sort the optimal order of transactions, maximizing returns both to the miner and to the pool. Under a new sorting logic, a specific transaction with a lower gas price could be more beneficial to the pool than another one with a higher gas price, as the new arrangement could lead to a greater number of transactions and higher due fees. The energy employed to optimize these transactions is named Proof-of-Efficiency (PoE), whose process forges the hash of an alien blockchain. Based on this forged hash, the novel design also employs improved technical solutions, such as the liquidity buffer and a novel protocol named Listrack (Listen and Track). Furthermore, the alien forged state allows other smart contracts located on the mainnet to confirm transactions without relying on third parties.

Finally, it is worth reiterating that interpool is a combination of a wrapping method and a liquidity pool, taking advantage of their merged synergy. In interpool, the novel design capitalizes on the twin deposits that are locked when the liquidity provider injects liquidity, turning the excess capital into collateral to mint a synthetic coin. This freshly minted coin is backed by risk management, which can either trigger more collateral or eventually a liquidation. The ultimate liquidity that enforces reliability is provided by the solid guarantee given by design that there is always a liquidity provider to transfer alien coins whenever burning is required. The burning procedure is monitored by the Listrack protocol supported by the alien forged hash. At last, interpool's design is a standalone solution that can provide comprehensive interoperability for other smart contracts on the mainnet, which can benefit from the Listrack protocol and forged alien hash to enable atomic swaps.

The remainder of the paper is structured as follows. In the next section, entitled Methodology, the article discusses the rationale of the design. Subsequently, the main ideas are restated in the Conclusion section, followed by a Glossary and the References section. After the References section, there is a statement about the conflicts of interest, followed by an Appendix containing the interpool entropy deduction.

## 2. METHODOLOGY

### 2.1. Combining a liquidity pool and a wrapping method

As previously discussed, the main insight in interpool is to merge a liquidity pool and a wrapping method, whose combined synergy provides a cross-chain bridge. This combination takes advantage of several synergies and opportunities, including leveraging the capital injected by liquidity providers and optimizing the order book.

In typical liquidity pools, liquidity providers must inject a pair of tokens comprising wrapped coins and native coins. However, the latest versions of typical pools are only employing wrapped tokens, even for the native coins. On the other hand, interpool does not require the deposit of alien wrapped coins, which should instead be exchanged for collateral in native coins. Therefore, under this approach, one of the deposits, which is the one that corresponds to the alien coin, is the collateral for a synthetic asset to be minted inside interpool. However, it is important to note that the deposited collateral is only fit for use because part of the freshly minted token always belongs to the liquidity provider as part of the locked capital. This logic is better explained with a numerical example in Section 2.5.

The risk related to the freshly minted token is managed inside interpool, which contains all required data to mitigate market risk without relying on oracles to provide either external logs or the exchange ratio. Under the new design, the exchange ratio comes directly from the ratio inside interpool, while external events are monitored by Listrack. Before explaining the logic regarding the minting and burning of the synthetic asset, the article explains the alien hash forgery and the Listrack protocol in the following sections.

## 2.2. Forging the hash of an alien chain inside interpool

Acting as a cross-bridge between two cryptocurrencies, interpool is required to listen to the alien chain in the mainnet. The first question is how to write such a listened alien state inside interpool. Unlike other approaches in the industry that must confirm every transaction outside the mainnet by a set of nodes acting at that alien chain, the novel protocol (Listrack) instead requires:

A. The raw transaction signed.

B. The *SPV* (simple payment verification (Nakamoto (2008)) of the raw transaction

C. The forged alien chain.

The approach described above is the same as a light client embedded in the mainnet. The Listrack protocol, which is explained in detail in the next section, is based on comparing the forged alien hash to the hashed tree of the SPV that also contains the hashed transaction. Under this approach, an alien transaction can be executed by a single party, rather than depending on the signatures of multiple parties using a threshold signature.

Before confirming a transaction using Listrack, the first step is to forge the hash of the alien blockchain. The idea behind this forgery is that misbehaviors, such as front running, could be converted into an optimization process to write an alien hash on the mainnet. As discussed before, front running is a misconduct tactic supported by the fact that transactions are executed by the miner based on their respective gas prices. Rather than sorting transactions based on their gas prices, sorting could be optimized to enable the maximum number of transactions. In this scenario, miners would benefit from more transactions being added to the block, and the optimizer would receive fees due to its computational effort. Furthermore, the users can also benefit from this approach because more transactions are processed, leading to more liquidity and efficiency. It is worth noting that the transactions subject to optimization can be either regular transactions (exchange) or liquidity injections supplied by liquidity providers.

The question that arises is how the number of transactions could increase in this optimization procedure. A good example is that a single large transaction holding a great volume can disrupt the exchange ratio, eliminating some transactions from the round due to their respective price constraints. These constraints are both price and volume ranges configured by the user to either enable the transactions or not. Therefore, the correct order of transactions located in the mempool is a mathematical puzzle whose solution can maximize the volume of transactions. For ease of reference, the optimizer is referred to as the booster[1] in the remainder of the article.

One might question whether the optimization procedure could be performed by the miner itself. This possibility indeed exists, enabling regular miners to participate in the optimization race. Such a race is accessible to anyone with access to the mempool and a public key on the mainnet to receive rewards. In this race, it is essential to estimate the trade-off between waiting for new transactions in the mempool and the time required for their optimization. However, a significant issue remains: how will the alien hash be forged in the interpool?

---

[1] see glossary for usual terms used in this article

The forgery of the alien hash should be embedded in the puzzle's solution, which can be acknowledged as a Proof-of-Efficiency (PoE) procedure. This forgery is described by the following rationale:

A. Every raw transaction in the mempool must be hashed and stored locally for optimization.

B. The booster runs a full client of the alien chain and monitors the production of alien blocks. However, rather than forging the last block hash, the booster must forge the finality hash, which is the hash of the block that corresponds to the last block minus the finality period of the alien chain. Using Bitcoin as an example, the booster must forge the Bitcoin hash that corresponds to the one located six blocks ago – for Bitcoin, it is usual to evaluate finality in 6 blocks. However, other applications may consider finality only after 24 hours (approximately 144 blocks). Thus, finality is a parameter that should be set in common agreement with participants in interpool. However, for other alien chains that use Proof-of-Stake, such as Ethereum, finality is not an issue anymore due to its deterministic consensus.

C. Based on its own batch of transactions retrieved from the mempool, the booster must run its optimization algorithm. The algorithm must have the following two goals, ordered by priority:

   a. Present a maximum return to the miner based on the total gas consumed by the batch of transactions. Therefore, the total gas consumed is the benchmark used by the miner to choose the best booster. This is commonly known as MEV (maximum extractable value) that a miner can receive in a batch of transactions. Without the proper optimization provided by boosters, the regular order of transactions based on gas price would collect lower fees than the optimized solution.

   b. Present a large volume of transactions. Under this assumption, the larger the volume of transactions, the greater the amount of fees the booster will receive inside the interpool. The booster will receive intertoken, which is the name of the minted token based on the alien coin. Choosing to pay the booster in intertokens instead of native coins encourages the booster to adhere to honest behavior. If a booster is a malicious actor who does not forge the correct alien hash, this behavior will harm his interests in redeeming intertoken for alien coin because the interpool will malfunction.

Therefore, the optimization algorithm is an objective function that aims to maximize, at first, gas returns to the miner and, second, a greater volume of transactions. However, there are still two constraints that are essential to the logic of the design: forgery of the alien hash and payment certainty to the booster. These constraints must be embedded in optimization, following the rationale described in the next steps and shown in Figure 2.

In the forgery scheme seen in Figure 2, the first bit of the optimized transaction must contain the first bit of the alien hash, while the last bit of the optimized transaction must present the first bit of the booster public key. Finally, the alien hash is forged by joining in a set the first bit of every 256 initial transactions. Likewise, the public key of the booster can be retrieved by joining in a set the last bit of every 256 initial transactions.

| #   | Optimized Transactions in Order (given by their hash in 256 bits) | | |
| --- | --- | --- | --- |
|     | first bit \| alien hash | middle 254 bits \| random selection | last bit \| booster public key |
| 1   | 0 | 00100101... (254 bits) | 0 |
| 2   | 1 | 11000011... | 1 |
| 3   | 0 | 00111100... | 0 |
| 4   | 1 | 10101001... | 0 |
| 5   | 0 | 01001110... | 1 |
| ... | ... | | |
| 256 | 0 | 00111000... | 1 |
| 257 | 10100111101001111010011110100111101001111010011110100111... | | |

Figure 2: Optimized schema of hashed transactions. The first bit in each of the 256 initial transactions is appended to a set that becomes the alien hash. Likewise, the last bit in each of the 256 initial transactions is appended to a set that becomes the booster public key.

Despite the advantages, the design is not deterministic. If the booster public key were not required, the same puzzle would be posed to every competitor, resulting in a likely deterministic enigma; however, even in these circumstances, the puzzle would not be entirely deterministic because each booster can gather a varying number of transactions in their batch. Furthermore, another stochastic component is that each booster must decide whether to add or remove transactions that have the same sender and nonce. In most cases, the wisest decision should be to consider only the one with the highest gas price because other pools and environments are out of the booster control. Furthermore, it is important to note that even the liquidity providers' transactions that inject liquidity into the interpool can be subject to optimization. Finally, it seems that such a semi-deterministic challenge poses an interesting puzzle because the chance of a booster being chosen relies on both computing power and randomness. This approach leaves some chance for smaller boosters to face major computing powers.

It is worth discussing entropy with regard to optimization. However, even before reaching entropy, the basic assumptions must be stated. In the schema shown in Figure 2, one of the assumptions is that the pool has more than 256 transactions. If the quantity is below this threshold, a viable alternative is to lock the first two and the last two digits of each optimized transaction, decreasing the number of transactions down to 128. On the other hand, increasing the number of locked digits in each optimized transaction reduces possible optimized solutions due to the lower degrees of freedom. Under these assumptions, a derived formula for computing entropy is shown in Equation (1) (deduced in the Appendix):

$$H(n, k, hash_{bits}) = \log\left[\left[\frac{\left(\frac{n}{k}\right)!}{\left(\frac{n - hash_{bits}}{k}\right)!}\right]^k \cdot (n - hash_{bits})!\right] \quad (1)$$

where $n$ is the number of transactions in the batch.

$hash_{bits}$ is the length of either the forged alien hash or the booster public key.

$k$ is 2 power the locked digits in a single transaction. $k = 2^{locked\ digits}$.

The locked digits are employed to either forge an alien hash or to forge a booster public key. In the example shown in Figure 2, there are two locked digits to forge the alien hash and the public key; therefore, $k = 2^{locked\ digits} => k = 4$.

It is important to notice that when equation (1) does not present constraints ($locked\ digits = 0\ and\ hash_{bits} = 0$), entropy would reach $\log(n!)$, which is the log of the possible permutations. The same equation shows that when the number of locked digits ($k$) increases, the number of possible combinations is lower, decreasing entropy. Another insight from the same equation is related to the number of $hash_{bits}$: the greater this number, the lower the entropy of the system.

Furthermore, the number of locked digits can be switched to either increase or decrease difficulty for boosters. Additionally, if one blockchain has a block production slower than its counterparty alien blockchain, other locked digits may be required to forge multiple alien states in a single block.

In the event that one booster does not provide the correct hash of the alien block, the next block must contain both the missed alien hash and the current one. In this case of failure, the number of locked digits would be increased to store the missed hashes and the current one. One question that can arise is whether it is possible to perform a cross-check on a potentially malicious forged hash. This cross-check in interpool turns out to be feasible because interpool's smart contract receives multiple alien hashes from liquidity providers proving their burning transfer. The burning transfer, which is discussed in the next section, contains a signed transaction and its SPV. As the SPV presents the hash of the alien tree, it is possible to compare multiple hashes from liquidity providers with the one provided by the booster.

Finally, interpool only becomes economically viable when the number of transactions in each block reaches a minimum threshold. This minimum transaction threshold proves to be an issue in initiating interpool because the number of transactions will inevitably be low in the beginning. A viable solution is to allow, for a short time, the use of other external wrapped tokens until a minimum volume threshold is met. As soon as the minimum volume is achieved, the short-lived external tokens are gradually removed from interpool, leaving only intertokens. In this scenario, minting intertokens begins only when the pool reaches this minimum threshold.

### 2.3. Listrack Protocol

Besides the forged alien hash, another required tool in interpool is the Listrack Protocol. These two tools enable interpool to be a MEB pool, where minting, exchanging, and burning occur in the same place. The Listrack Protocol was conceived to be a smart contract that can be used for any atomic swap between the mainnet and an alien chain, without requiring oracles or threshold wallets. While Listrack in mainnet is a sole smart contract, the Listrack in interpool is rather part of the smart contract that governs interpool. This Listrack logic coded inside interpool supports intertoken burning when users put in a claim to redeem alien tokens. The Listrack logic is described in the following steps for either a burning procedure in interpool or a plain vanilla swap, in which the latter is illustrated in Figure 3.

### A. Transaction Agreement
*Plain Vanilla Swap*
Two parties, namely Mike (mainnet) and Alice (alien chain), agree on the terms of a transaction. In this agreement, mandatory fields must be provided, such as public keys from both parties (mainnet and alien chain) and the transaction details regarding the exchange ratio and volume. Mike and Alice must present enough funds in either native coins or alien coins, respectively, for their part of the agreement. Furthermore, Alice must hold collateral in native coins; if she does not transfer the alien coins, this collateral will be slashed as a penalty for her inaction.

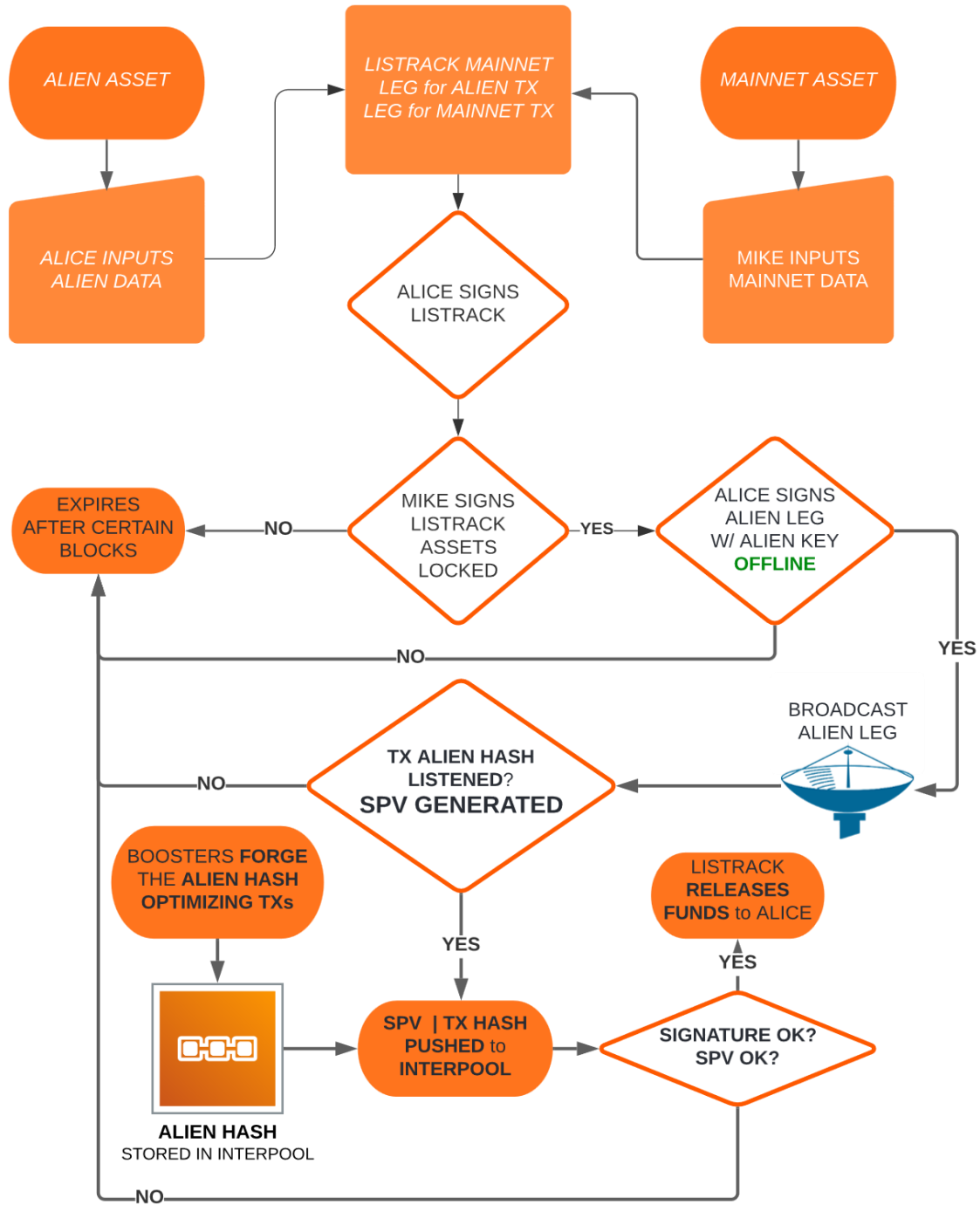

Figure 3: The Listrack protocol for a plain vanilla swap

*Interpool*
In Interpool, one of the parties is the liquidity provider, and the other one is the user claiming alien coins by burning intertokens. Their public keys on mainnet and the alien chain are already stored in Interpool's smart contract. However, these parties can change their public keys at any moment by pushing data to Interpool's smart contract.

### B. Mainnet signature
*Plain Vanilla Swap*

Both parties sign the agreement on the mainnet. Mike's funds and Alice's collateral are locked until Listrack listens for and tracks Alice's transfer on the alien chain.

*Interpool*

In interpool, the collateral of the liquidity provider was locked when he joined interpool.

### C. Signing offline
*Plain Vanilla Swap*

Alice signs the alien transaction offline using her favorite front-end application or wallet provided by her host.

*Interpool*

In interpool, the liquidity provider also signs the burning transaction requested by interpool's smart contract. The liquidity provider uses a front-end application that receives alerts from Interpool requiring first to confirm his public key and, second, the burning transfer.

### D. Pushing SPV and signed transaction
*Plain Vanilla Swap*

As soon as Alice's transfer to Mike is confirmed in the alien chain, her front-end application stores the SPV path along with Alice's signature. The SPV, as previously discussed, is an acronym for Simple Payment Verification (Nakamoto, 2008), which contains the transaction path from the first leaf up to the root hash of the Merkle Tree. Finally, the signed raw transaction along with the SPV is pushed to Listrack in the mainnet.

*Interpool*

In interpool, the liquidity provider acts as Alice, signing the raw transaction. After the transaction is confirmed on the alien chain, the liquidity provider sends the SPV along with the signed raw transaction. Similar to Alice's case, the SPV contains the hash of the alien chain that is used for comparison against the forged hash provided by the booster. As previously mentioned, it is important to note that the SPV also contains the alien hash header, which supports the comparison against the one forged by the booster.

### E. Logical operations on SPV and signed transaction
*Plain Vanilla Swap*

When Listrack receives Alice's message, the smart contract performs two logical operations:

a. It verifies Alice's signature against her public key stored in Listrack.
b. It verifies whether the hash of the transaction, hashed several times following the SPV path, is equal to the alien forged hash.
   If both logical comparisons are successful, Listrack finally releases Mike's funds to Alice, terminating the operation. If a certain period elapses without the confirmation, the transaction fails. In this case, Mike's funds are released, and Alice's collateral is slashed

*Interpool*

The final step in interpool follows the logic described for a plain vanilla swap. After receiving the message from the liquidity provider, interpool's smart contract makes two logical procedures, as discussed before: it verifies the signature over the transaction and compares the SPV path with the alien forged hash. If verifications are successful, interpool's smart contract releases the corresponding collateral to the liquidity provider. In case of failure, the liquidity provider's collateral is slashed, and other subsequent procedures are further described in the following sections.

Finally, the time variable must be discussed due to possible failures in Listrack, as shown in Figure 3. As explained before, the Listrack logic can be coded in either a sole smart contract or inside Interpool, in which both cases can be triggered by several distinct users. Every time a transaction is initiated in the Listrack logic, it runs a boolean comparison to verify whether the transaction is the first one of a new block. If a new block begins with this pending transaction, the Listrack logic runs some procedures before executing the first transaction. Therefore, there are some procedures that Listrack logic runs at the beginning of every new block, such as:

A. In interpool, Listrack logic runs tailored algorithms for risk management that release funds, slash collateral, liquidate positions, and estimate the level of the liquidity buffer (yet to be discussed).
B. In a plain vanilla swap, Listrack logic can eventually slash collateral (e.g., Alice's collateral) and release locked funds (e.g., Mike's funds).

## 2.4. Basic Assumptions in Interpool

Before explaining the MEB process in detail, the basic assumptions of the design are stated below:

A. Interpool is an exchange pool between a native coin on mainnet and an alien coin that is wrapped inside interpool, in which the latter effectively becomes the ERC-20 token named intertoken. In the current design, it is not possible to exchange two different alien coins in the same interpool; however, two intertokens minted in their respective interpools can be regularly exchanged in another liquidity pool.
B. Intertoken is entirely managed inside the smart contract of interpool. This design is only possible due to the support given by the forged alien hash and the novel Listrack protocol, discussed in previous sections. Intertoken has a 1:1 parity ratio, which means that burning 1 intertoken allows a claim to redeem 1 alien coin.
C. Interpool pool runs under the constant product approach, which is a type of automated market maker mechanism commonly used in decentralized finance, as shown in (2):

$$intertoken_{volume} * native_{volume} = constant \qquad (2)$$

D. The liquidity provider places just one deposit in native coins when he joins the interpool. His deposit is split into two twin deposits that hold the same number of native coins. One of the twin deposits backs intertoken minting, with the exact volume corresponding to the exchange rate between native tokens and intertokens (refer to Figure 4 and 5). The product of the freshly minted intertokens and the remaining twin deposit determines the interpool coins due to the liquidity provider, which are his share in the interpool. Regardless of market movements, the interpool coins held by the liquidity provider remain constant, but such coins are locked until he transfers alien coins due to the equivalent intertoken burning claim. Even though interpool coins are locked until burning, the liquidity provider can withdraw his deposit at any moment, in which case his due balance is refunded.

On the other hand, the liquidity provider's share depends on his amount of interpool coins compared to the overall locked capital in the interpool, with such a share being the mechanism to give him the due fees.

E.  As discussed in the previous topic, interpool coins belonged to each liquidity provider become an ERC-20 token as soon as their burning commitments are met. Therefore, as soon as the burning constraint is released, interpool coins can be swapped for any other token or native coin in the mainnet.

F.  At the beginning of each block, as mentioned in the previous section, interpool runs tailored algorithms for risk management that slash collaterals, liquidate positions, release funds, and estimate the level of the liquidity buffer.

G.  The paper does not cover how to bootstrap interpool; however, it provides one possible alternative: interpool can begin using other wrapped tokens to ignite the pool until it reaches the minimum threshold for minting intertokens.

H.  The magnitude of the fees due to the boosters and the liquidity providers must be subject to further research in a PoC (proof-of-concept) environment. The article suggests that these fees should be variable in accordance with the liquidity inside the interpool.

I.  Governance rules are not covered in this article.

J.  Misbehaviors are mentioned in this article without specifying the penalty for malicious actors.

K.  A front-end application must be developed to assist the liquidity provider in monitoring various logs, such as:

   a. Receive logs to stake more collateral. Furthermore, the front-end application can also request the liquidity provider to sign the transaction.

   b. Receive logs flagging that his burning role to transfer alien coins is about to happen, requiring the push of a new public key whether such key has changed since he provided liquidity to the interpool.

   c. Receive a log requiring an immediate alien coin transfer to a user, whose public key is included in the log.

   d. Receive logs informing him of his balance in interpool.

## 2.5. Minting, Exchanging and Burning (MEB) in interpool

The best way to explain the design of Interpool is to first take the perspective of a liquidity provider, and then, in a second step, explain the design from the user's perspective. From the liquidity provider's point of view, interpool logic is shown in Figure 4 and described in the following sections.

### A. Liquidity provider joins interpool

The liquidity provider joins interpool knowing the exchange ratio between intertoken and the native token. In a hypothetical numerical example described in Figure 5, this exchange ratio is 1:2.5, which states that 1 intertoken is equal to 2.5 native tokens.

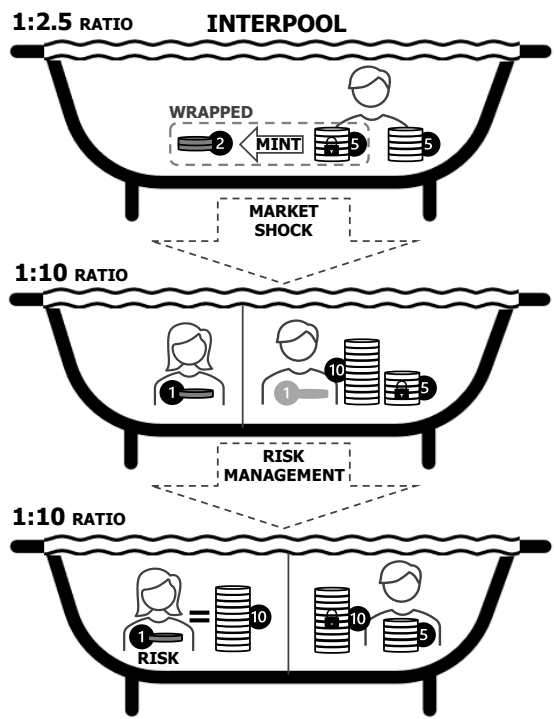

Figure 4: Interpool design hit by an extreme market shock (refer to section 2.6 and Figure 5). At the initial stage, interpool presents a 1:2.5 ratio exchange between intertoken and the native coin, resulting in minting 2 intertokens backed by 5 native coins and a twin deposit of 5 native coins, comprising 10 native coins injected by the liquidity provider. At market shock, the exchange ratio jumps to 1:10 ratio, which leads to the liquidity provider holding 10 native coins and 1 intertoken due to the constant product formula, while the other 1 intertoken now belongs to Alice. However, the 1 intertoken held by the liquidity provider is a virtual coin (gray) that only becomes real to him when burning claims are met. Due to the market shock, the risk relies on the 1 intertoken held by Alice that must be backed by 10 native coins, rather than the collateral comprised of 5 native coins; therefore, the liquidity provider must inject more 5 native coins to be used as collateral, otherwise his position may be liquidated.

|  | exchange ratio | constant product liquidity pools | intertoken | native coins |
|---|---|---|---|---|
| **time_0 \| joining interpool** | 1:2.5 | 10 | 2.00 | 5.00 |
| native coin deposit |  |  | 5.00 | 5.00 |
|  |  |  |  |  |
| **time_1 \| moderate market shock** | 1:5 | 10 | 1.41 | 7.07 |
| intertoken subject to risk |  |  | 0.59 |  |
| risky intertoken in native coins |  |  |  | 2.93 |
| native coin collateral |  |  |  | 5.00 |
| collateral remaining |  |  |  | **2.07** |
| total balance |  |  |  | **9.14** |
|  |  |  |  |  |
| **time_2 \| extreme market shock** | 1:10 | 10 | 1.00 | 10.00 |
| intertoken subject to risk |  |  | 1.00 |  |
| risky intertoken in native coins |  |  |  | 10.00 |
| native coin collateral |  |  |  | 5.00 |
| collateral remaining |  |  |  | **-5.00** |
| total refund due to liquidation |  |  |  | **5.00** |

Figure 5: risk management enabled by collateral in native coins.

B. **Liquidity provider only stakes native coins**

Rather than providing two deposits comprising 2 wrapped coins and 5 native coins (1:2.5), which is the usual procedure in liquidity pools, the liquidity provider stakes 10 native coins in the mainnet. Therefore, there are 5 native coins remaining to back 2 wrapped alien coins, which becomes the collateral to mint 2 intertokens.

C. **Minting intertoken**

As mentioned before, as soon as the liquidity provider stakes 10 native coins, 5 coins are locked to back the minting of 2 intertokens. These 5 coins that are locked to be used as collateral do not take part in the liquidity pool. On the other hand, the remaining 5 native coins, along with the freshly minted 2 intertokens, are added to the pool for regular exchange.

D. **Moderate market shock**

In a hypothetical next block, due to high volatility, the exchange ratio moves to 1:5. In this new ratio, based on the constant product formula, 2 intertokens of the liquidity provider turns into 1.41 intertokens and its native coin goes to 7.07. At this moment, 0.59 intertokens are out of the interpool in the mainnet, held by third parties. Thus, these 0.59 intertokens, subject to market risk, must be backed by the interpool. In this new exchange ratio, 0.59 intertokens corresponds to 2.93 native coins, getting closer to the limit of collateral detained by the liquidity provider, which is 5 native coins.

*Liquidity provider chooses to withdraw his coins*

If the liquidity provider chooses to withdraw his coins in interpool, he would receive a net value of 9.14 native coins plus the fees earned by his supply service. In this case, the 2.93 native coins due to the interpool are added to the liquidity buffer, which simultaneously burns 0.59 intertokens in the same buffer. Burning 0.59 intertokens is required because the commitment to transfer 0.59 alien coins is over, due to the exit of the liquidity provider. Moreover, there is no more collateral to back these 0.59 intertokens, which must be burned in the liquidity buffer.

Without the liquidity buffer to assist in burning during liquidations, interpool would have to exchange 2.93 native coins for 0.59 intertokens in the exchange environment, artificially draining the intertoken inventory. Using the liquidity buffer, several successive liquidations can be absorbed without leading to higher volatility. It is worth noting that the liquidity buffer is a small pool of native tokens and intertokens, which are not subject to exchange. Fees and penalties are stored inside the liquidity buffer, whose inventory is used for paying boosters and liquidity providers.

*Liquidity provider chooses to remain in interpool*

On the other hand, the liquidity provider could have chosen to stay in the interpool, in which case he would supply 2.93 native coins to reinstate the risk balance. In this case, the interpool would have flagged the liquidity provider through a log captured by his front-end application, requiring him to supply more tokens.

It is important to notice that this high-shock event has a low probability of happening from one block to another.

### E. Burning intertoken triggers an alien coin transfer

Even though the liquidity provider does not have the immediate obligation to transfer alien tokens when he joins interpool, this pending transaction will inevitably happen at some moment. The exact moment that the liquidity provider transfers alien tokens is when interpool emits a log for the liquidity provider, flagging that a burning request must be met by him. Therefore, from the first moment, the liquidity provider has been aware of the promise that he would eventually be requested to transfer alien tokens to another wallet.

Moreover, as the moment of burning approaches, the penalty for the liquidity provider increases if he leaves interpool. The penalty applied to the liquidity provider can be based on a suitable decay function starting at the burning commitment, such as using a Dutch auction function. However, the burning commitment cycle depends on the burning appetite of interpool, which is regularly updated and broadcasted to the liquidity providers.

The order in which liquidity providers are requested to make a transfer depends on a rotating set of incoming liquidity providers. In other words, a liquidity provider that joined the pool 50 blocks ago will be requested to make a transfer earlier than one that joined the pool 10 blocks ago. Based on the rotating set and burning appetite, interpool will emit events for certain liquidity providers, flagging them as the next group for providing alien liquidity. When these liquidity providers joined interpool, they uploaded their public key that contains the alien coins. If their alien wallet is gone for any reason, a liquidity provider has the chance to change their public key one block before his transfer without incurring any penalty. In order to increase reliability, there could be incentives to create a new category of liquidity providers: the full providers.

The full providers should be incentivized for keeping a buffer of alien coins immediately available for any burning request. Therefore, if a burning request pops up in the interpool, full providers have the priority to fulfill the transfer in the next block. This approach enforces the security and reliability of the interpool in providing liquidity.

On the other hand, the regular providers are the liquidity providers who wait until the end of each burning cycle to make their alien coin transfer. This approach enforces liquidity but poses a new type of risk, which is leaving some regular providers without ever providing liquidity because full providers always take the lead. This never-liquidity issue for some regular providers can be mitigated by flagging a time limit for their alien coin transfer, regardless of whether there is plenty of liquidity in the interpool.

It is worth remembering that when a liquidity provider finally transfers alien coins to comply with their minting obligation, their collateral is refunded, and they receive an interpool coin that can be exchanged as an ERC-20 token. This interpool coin is the stock share of the interpool that becomes available for exchange as soon as the burning liability is over.

Finally, it is important to note that the burning design is only possible due to the foundations provided by the interpool architecture, which include the forgery of the alien hash and the Listrack protocol. Without these foundations, interpool would not be able to detect whether an alien transfer occurred or not, and thus would miss releasing collateral or imposing penalties.

## 2.6. Extreme market shock

In a more unlikely event, the exchange ratio could have tapped 1:10. In this new ratio, based on the constant product formula, 2 intertokens of the liquidity provider turn into 1 intertoken, and his position in native coins goes to 10 (refer to Figures 4 and 5). Unlike the other scenario, the 5 native coins held as collateral are not enough to back 1 intertoken at risk, which actually corresponds to 10 native coins. Thus, as the collateral is not enough to back the intertoken at risk, part of the liquidity provider balance in native coins is slashed, leaving a net balance of 5 native coins. Therefore, even in extreme events, the design can support a market shock. However, interpool governance may request the liquidity provider to inject 5 more native coins to reinstate collateral; otherwise, his position is going to be terminated,

Beyond collateral requests to the liquidity providers, each market shock triggers appropriate actions that depend on interpool governance and the extent of the shock. Some of these actions, properly designed to mitigate risk, are shortlisted below:

A. More collateral than the exact amount required to back the exposed risk may be requested due to high volatility.
B. Regardless of whether more collateral is injected or not, a position can inevitably be liquidated.
C. The liquidity buffer within interpool can be activated to absorb volatility, which is discussed in detail in the following sections.

## 2.7. Users in Interpool

The interpool guidelines for users are similar to those that already exist in typical liquidity pools. In interpool, users who intend to exchange either intertokens or native coins can interact with the smart contract without restrictions, except for the constraints that users themselves establish. In typical liquidity pools, the pair of tokens are commonly wrapped versions of foreign coins, whose wrapping method can pose several unpredictable risks (as discussed before). However, interpool significantly enhances trust because it does not depend on external wrapping methods. The wrapped alien coin in interpool, named intertoken, is entirely managed by interpool.

Moreover, as wrapping occurs inside interpool, it avoids typical costs related to wrapping and burning. On the other hand, boosters are new actors that require extra fees for their service regarding optimization, which could lead to more costs for users. Despite the new fees for boosters, optimization surpasses the negative effect by enlarging liquidity due to a greater number of transactions. As liquidity grows through optimization, the fees charged for each transaction can significantly decrease.

## 2.8. The Liquidity Buffer

Even though the liquidity buffer has already been introduced, the main foundations and its functioning mechanism are yet to be discussed. The liquidity buffer comprises two stacks:

A. A stack of native coins successively generated by slashing collaterals, receiving penalties, or collected fees due to transactions.
B. A stack of intertokens piled up by collecting fees from transactions.

Furthermore, the demand for each coin in the liquidity buffer arises from the following requests:
A. Intertoken is used for boosters' payments, failed burning operations, and early liquidations.
B. Native coin is the currency used to pay fees to the liquidity providers.

As mentioned before, the liquidity buffer is not part of the exchange pool; it is kept apart in another layer of interpool, in which the levels of each coin do not necessarily respect the constant product formula.

It is worth discussing the positions that are liquidated in interpool due to market shocks, early withdrawals, or failed burning transfers. In such cases, there could be exposed risk in intertokens (refer to Figures 4 and 5), which requires burning intertokens as soon as the collateral is received. The need to burn intertokens due to such abnormal operations arises from the broken commitment to transfer alien coins that are not supported anymore by their respective collateral.

One solution for burning intertoken due to abnormal operations is to purchase the required amount directly in interpool; however, purchasing intertoken could change the parity ratio, leading to an artificial price movement that could generate a snowball effect. In order to solve this issue, intertoken is burned inside the intertoken stack in the liquidity buffer, as soon as the collateral is received. This action does not disrupt interpool, because this mechanism is set apart in the liquidity buffer.

Finally, the liquidity buffer can be used to mitigate volatility. Based on a tokenomics algorithm, which is integrated with the one designed for evaluating fees, a flag can be triggered whenever excess volatility is detected. In this case, the flag ignites a subsequent event that can deploy part of the liquidity buffer into the interpool, mitigating volatility.

## 2.9. Fees in Interpool

Liquidity providers earn yields based on their share of interpool, similar to regular liquidity pools in decentralized finance. These fees commonly range from 0.05% to 3% for each transaction and must be extensively discussed for proper functioning in interpool. Rather than making the fee a fixed constant, a proper algorithm can be used to calibrate liquidity in interpool using variable fees. Under this system, when liquidity in interpool reaches a minimum threshold, the fee can increase to attract more liquidity providers. Conversely, the fee can decrease when there is plenty of liquidity.

It is worth remembering that the liquidity buffer is comprised of two stacks: one is made up of intertokens that are stacked successively by earning fees in intertokens; and the other stack is comprised of native tokens earned by fees and penalties. Even though the liquidity buffer is composed of native tokens and intertokens, liquidity providers always receive their due fees in native tokens, while boosters receive their due fees in intertokens. The fee collection method can be switched from charging transactions in native tokens to intertokens or vice-versa, depending on the overall liquidity.

Switching the fee collection method between native coins and intertokens will depend on an algorithm that calibrates the correct threshold level in the liquidity buffer for both stacks. This algorithm should evaluate volatility over time, estimating critical thresholds for both intertokens and native coins. As mentioned before, the proper fee percentage can also be derived from this tailored algorithm, which is subject to further research. Under normal circumstances, there is a sustained demand for collecting fees through both methods, as follows:

A. There is a steady demand for intertokens due to failed burning transactions, liquidations, and boosters' payments. This demand can be met, for example, by charging a fee of 1% over a transaction of 100 intertokens. In this 1% fee case, interpool collects 101.01 intertokens, which generates a 1.01 intertoken fee. As a result, such fees are successively stacked in the liquidity buffer.
B. The native coin demand is constant because liquidity providers must receive their supply service in native coins, The liquidity buffer in native coins can either be stacked by switching fees to native coins or by receiving penalties and collaterals.

**3. CONCLUSION**

The potential of full interoperability between distinct blockchains presents significant opportunities for decentralized finance, yet existing solutions face notable challenges, especially the lack of a liquid atomic swap without third parties. The interpool design offers a comprehensive and novel solution that combines in a single place minting, exchanging, and burning (MEB). The MEB approach is founded on two important insights, as follows.

The first insight is to turn the front-running issue in the mempool into a solution to forge an alien hash. The alien hash, which is the header hash of a foreign chain, is used by a novel protocol named Listrack (Listen and Track) to verify alien transactions without relying on oracles or threshold account signatures. The alien hash is forged under a Proof-of-Efficiency algorithm (PoE) that optimizes transactions in the mempool by maximizing returns for both the miner and the booster (optimizer). During optimization, the booster writes the hash of the alien chain and its own public key in locked digits, which are forged in every optimized transaction.

The second insight is to use one of the twin deposits in a liquidity pool to back the minting of a wrapped asset. Unlike other liquidity pools, the locked capital is not only used for providing liquidity but also serves as collateral to support a secure wrapped coin. Supported by Listrack, such a wrapped asset is minted directly inside interpool without relying on third parties. With proper risk management in interpool, there is always sufficient collateral for a wrapped coin until the liquidity provider transfers the due amount in alien coins when a burning claim is set. The burning mechanism, founded on Listrack, gives interpool the capability to monitor whether an alien transfer has occurred, which otherwise results in a collateral slash. This reliable burning capability granted to interpool eternalizes liquidity without relying on third parties.

In its essence, interpool represents a significant advancement in decentralized interoperability solutions, combining the strengths of liquidity pools and wrapping methods into a single, efficient framework. Its capability to manage risks, optimize transactions, and maintain liquidity without dependency on centralized entities marks a pivotal step forward for decentralized finance. Future exploration into governance rules, fee dynamics, and a comprehensive Proof-of-Concept (PoC) will further refine and validate the interpool model as a robust solution for blockchain interoperability.

**GLOSSARY**

*actors*
    Liquidity providers, users and boosters that have a pair of public and private keys on both blockchains (alien chain and mainnet).

*alien blockchain*
    The blockchain whose asset will be wrapped in interpool. The alien blockchain can be either Turing complete or not Turing complete.

*alien coin*
    The coin that exists on the alien blockchain.

*booster*
    The individual who runs an optimization procedure over a batch of transactions subject to front-running, which can be regular transactions (exchange) or liquidity injections provided by liquidity providers. The optimization generates higher incomes for the miner and a greater number of transactions for interpool. The optimization procedure forges the hash of an alien chain using the optimized order of transactions. The optimization procedure also forges the public key of the booster, ensuring his payment.

*forged alien hash*
    The block header hash of an alien blockchain that is forged by a booster inside the interpool

*full provider*
    A liquidity provider that is capable of transferring alien coins whenever interpool flags a burning claim. Due to his premium service, the full provider receives more fees than a regular provider.

*interpool*
    A pool that is a combination of a wrapping system with a liquidity pool. In Interpool, unlike other liquidity pools, the wrapped asset is minted inside the pool. Interpool takes advantage of Listrack for securely burning wrapped assets in exchange for alien coins, which eternalizes liquidity under proper risk management.

*interpool coins*
    The interpool coins are the product of supplied intertokens and native coins at the moment a liquidity provider injects liquidity into the interpool. These interpool coins only become ERC-20 tokens when their holders, who are the liquidity providers, meet their respective burning obligations as logged by the interpool. The shares held by each liquidity provider in the interpool are weighted by the corresponding amount of coins each one holds.

*intertoken*
    The coin minted inside interpool that wraps an alien coin. The intertoken exchange ratio to its corresponding alien coin is 1:1.

*liquidity buffer*
    A safety pool set apart from the exchange environment, consisting of intertokens and native coins. The level of native coins and intertokens in the liquidity buffer presents quantities that do not necessarily correspond to the product constant

formula. The buffer is employed to burn intertokens in failed burning transactions and early liquidations. Moreover, the liquidity buffer can be used to manage volatility and coin shortages.

*liquidity provider*
    The individual who injects two twin deposits in native coins into the interpool. One of the deposits is blocked to be used as collateral for minting intertokens, while the other one goes directly to the interpool. Like other liquidity pools, the liquidity provider receives fees for their shares in the interpool.

*Listrack*
    The novel protocol designed for atomic swaps that do not rely on third parties nor oracles, which is supported by the forged alien hash inside interpool.

*mainnet*
    The Turing complete blockchain where interpool is located.

*native coin*
    The mainnet currency.

*Proof-of-Efficiency (PoE)*
    The process that solves an optimization puzzle, with the main objective of maximizing the number of possible transactions, which otherwise would be lower due to issues such as front-running. Therefore, optimization targets increasing transactions in the block by properly ordering their execution, enabling maximum fees for both the miner and the booster (optimizer). The batch of transactions subject to optimization can be either regular transactions (exchange) or liquidity injections supplied by liquidity providers.

*regular provider*
    A liquidity provider that only engages in transferring alien coins when strictly necessary. The moment the regular provider must transfer alien coins depends on the burning cycle and the pending row of liquidity providers without transfers.

*SPV*
    SPV, first mentioned in Nakamoto (2008), corresponds to simple payment verification. In this approach, a transaction can be validated if its hash, combined with the hashes of the Merkle tree it belongs to, is equal to the hash of the mined block.

*user*
    The individual who is interested in exchanging native coins for intertokens or vice versa. A user can submit a claim at any moment to burn intertokens, which triggers the transfer of the corresponding alien coins to his alien wallet.

*wrapping in interpool*
    The procedure employed to mint intertoken that uses a deposit in native coins as collateral. This collateral in native coins is the twin of another identical deposit that is not collateralized. Under this design, the freshly minted intertoken and the other twin deposit are thrown directly into the interpool.

## CONFLICT OF INTEREST STATEMENT

# APPENDIX A
# Entropy in Proof-of-Efficiency (PoE)

The possible number of permutations in the optimization procedure is discussed in detail in this appendix. It is worth mentioning that the possible transactions are either the regular exchanges or the liquidity injected by liquidity providers. As a first approach to compute permutations, one can figure out that transactions can be arranged by their locked bits, as shown in the Figure 6 below:

| # | Round | Optimized Transactions in Order (given by their hash in 256 bits) | | |
|---|---|---|---|---|
| | | first bit \| alien hash | middle 254 bits \| random selection | last bit \| booster public key |
| 1 | A | 0 | 00100101… (254 bits) | 0 |
| 2 | A | 1 | 11000011… | 1 |
| 3 | A | 0 | 00111100… | 0 |
| 4 | A | 1 | 10101001… | 0 |
| 5 | B | 0 | 01001110… | 1 |
| 6 | B | 0 | 00110000… | 1 |
| 7 | B | 1 | 11000011… | 1 |
| 8 | B | 0 | 00011001… | 0 |
| 9 | C | 1 | 11000110… | 1 |
| 10 | C | 0 | 00101010… | 0 |
| 11 | C | 0 | 00111000… | 1 |
| 12 | C | 1 | 11010111… | 0 |
| 13 | D | 0 | 00101101… | 0 |

Figure 6: Transactions in a batch arranged for computing the possible permutations

As shown in Figure 6, 2 locked bits correspond to 4 possible outcomes, which implies that 4 optimized transactions will likely remove 1 degree of freedom in optimization. Therefore, after 4 optimized transactions in Round A, there are $\frac{n}{k} - 1$ transactions left subject to optimization (where $n$ = number of transactions and $k = 2^{locked\ digits} => 4$). In the next round of 4 optimized transactions (Round B), there are $\frac{n}{k} - 2$ transactions left subject to optimization. This logic will continue until $\frac{n}{k} - 64$ transactions, which corresponds to $\frac{n}{k} - \frac{hash_{bits}}{k}$, which equals to $\frac{n - hash_{bits}}{k}$. Thus, the number of possible transactions equals:

$$\frac{n}{k} \cdot \left(\frac{n}{k} - 1\right) \cdot \left(\frac{n}{k} - 2\right) \cdot (\ldots) \cdot \left(\frac{n - hash_{bits}}{k}\right) = \frac{\frac{n}{k}!}{\frac{n - hash_{bits}}{k}!} \qquad (3)$$

Equation (3) happens 1 time for each element in the group of $k$ individuals. Therefore, the number of possible combinations in the group of 256 ($hash_{bits}$) individuals is:

$$\Omega\ tx_{first\ hash_{bits}} = \left[\frac{\left(\frac{n}{k}\right)!}{\left(\frac{n - hash_{bits}}{k}\right)!}\right]^k \qquad (4)$$

After achieving the length of the forged hash ($hash_{bits} = 256$), the remaining transactions do not have constraints. Therefore, the possible transactions after forging the alien hash are:

$$\Omega tx_{after\ hash_{bits}} = (n - hash_{bits})! \qquad (5)$$

Finally, merging (4) and (5) leads to equation (6):

$$\Omega\, tx = \left[\frac{\left(\frac{n}{k}\right)!}{\left(\frac{n - hash_{bits}}{k}\right)!}\right]^k \cdot (n - hash_{bits})! \quad (6)$$

As the probability of choosing the right transaction in optimization is the same, Shannon's entropy (Shannon (1948)) formula finally leads to equation (1):

$$H(n, k, hash_{bits}) = \log\left[\left[\frac{\left(\frac{n}{k}\right)!}{\left(\frac{n - hash_{bits}}{k}\right)!}\right]^k \cdot (n - hash_{bits})!\right] \quad (1)$$